Date: September 09, 2021

# Influences of ALD Al$_2$O$_3$ on the surface band-bending of c-plane, Ga-face GaN and the implication to GaN-collector npn heterojunction bipolar transistors

Jiarui Gong[1], Jisoo Kim[2], TienKhee Ng[3], Kuangye Lu[4], Donghyeok Kim[2], Jie Zhou[2], Dong Liu[2], Jeehwan Kim[4], Boon S. Ooi[3], and Zhenqiang Ma[2,a]

[1]*Department of Physics, University of Wisconsin-Madison, Madison, Wisconsin, 53706, USA*

[2]*Department of Electrical and Computer Engineering, University of Wisconsin-Madison, Madison, Wisconsin, 53706, USA*

[3]*Department of Electrical and Computer Engineering, King Abdullah University of Science and Technology, Thuwal 23955-6900, Saudi Arabia*

[4]*Department of Mechanical Engineering, Department of Materials Science and Engineering, Massachusetts Institute of Technology, Cambridge, Massachusetts, 02139, USA*

[a] Author to whom correspondence should be addressed. Electronic mail: mazq@engr.wisc.edu

# Abstract

Due to the lack of effective p-type doping in GaN and the adverse effects of surface band-bending of GaN on electron transport, developing practical GaN heterojunction bipolar transistors has been impossible. The recently demonstrated approach of grafting n-type GaN with p-type semiconductors, like Si and GaAs, by employing ultrathin (UO) $Al_2O_3$ at the interface of Si/GaN and GaAs/GaN, has shown the feasibility to overcome the poor p-type doping challenge of GaN by providing epitaxy-like interface quality. However, the surface band-bending of GaN that could be influenced by the UO $Al_2O_3$ has been unknown. In this work, the band-bending of c-plane, Ga-face GaN with UO $Al_2O_3$ deposition at the surface of GaN was studied using X-ray photoelectron spectroscopy (XPS). The study shows that the UO $Al_2O_3$ can help in suppressing the upward band-bending of the c-plane, Ga-face GaN with a monotonic reduction trend of the upward band-bending energy from 0.48 eV down to 0.12 eV as the number of UO $Al_2O_3$ deposition cycles is increased from 0 to 20 cycles. The study further shows that the band-bending can be mostly recovered after removing the $Al_2O_3$ layer, concurring that the change in the density of fixed charge at the GaN surface caused by UO $Al_2O_3$ is the main reason for the surface band-bending modulation. The potential implication of the surface band-bending results of AlGaAs/GaAs/GaN npn heterojunction bipolar transistor (HBT) was preliminarily studied via Silvaco® simulations.

**Introduction**

For its high breakdown electrical field (4.9 MV/cm) [1] and high electron saturation velocity ($1.4 \times 10^7$ cm/s) [2], gallium nitride (GaN) has been the semiconductor material enabling the highest-performance solid-state radiofrequency (RF) electronic device at present [3]. Although GaN-based high electron mobility transistors (HEMT) have been well-developed over the last two to three decades since its first demonstration [4], the ultimate electronic potential of GaN for both RF and power electronic applications may be eventually exploited from bipolar junction transistors (BJTs) [5]. With a vertical three-dimensional current flow from emitter to collector and the exponential dependence of the collector current density on the bias, BJTs based on GaN are theoretically expected to provide a much higher current density than that of the GaN-based HEMTs. However, limited by the well-known poor p-type doping in GaN, the most critical challenge, implementing practical GaN-based BJTs has been impossible to date [6]. In additional to this major challenge, the preliminary investigations of GaN-based BJTs revealed that the surface energy band-bending of GaN needs to be taken into crucial consideration for GaN-collector bipolar transistor applications [7, 8].

Considering the fundamental physics limit of realizing effective p-type doping in GaN, an alternative approach to tackle the p-type doping challenge in GaN-based BJTs development is to replace p-type nitride materials with other p-type semiconductor materials, such as silicon (Si), germanium (Ge), indium phosphide (InP), gallium arsenide (GaAs), etc. and employ these p-type materials as the base region of GaN-based bipolar transistors. Nevertheless, the lattice mismatch between these p-type materials and GaN prevented the epitaxial growth of GaN [9, 10] and wafer bonding/fusion approach [7, 8] from producing an abrupt junction/interface between GaN and the p-type semiconductors with a low interface density of states. The recently demonstrated semiconductor grafting approach [11], which, in terms of the working mechanisms, fundamentally distinguishes itself from the heteroepitaxy and the wafer bonding/fusion approaches, has solved the lattice-mismatch challenge. The insertion of an ultrathin oxide (UO) (e.g.,

Al$_2$O$_3$) layer between GaN and semiconductors with good p-type doping provides an abrupt junction with a very low interface density of states, manifested by the ideal p-n junction diodes characteristics. The p-n junctions formed by the grafting approach, featured with a p-n diode ideality factor (*n*) of ~ 1.10, ultra-low reverse-bias current density, record-high current on/off ratio values, avalanche breakdown characteristics, etc. rival those of the epitaxially grown, lattice-matched p-n junctions. Following the grafting approach [11], record-breaking III-nitride devices [12-14] and preliminary pnp AlGaAs/GaAs/Diamond HBT [15] have been reported.

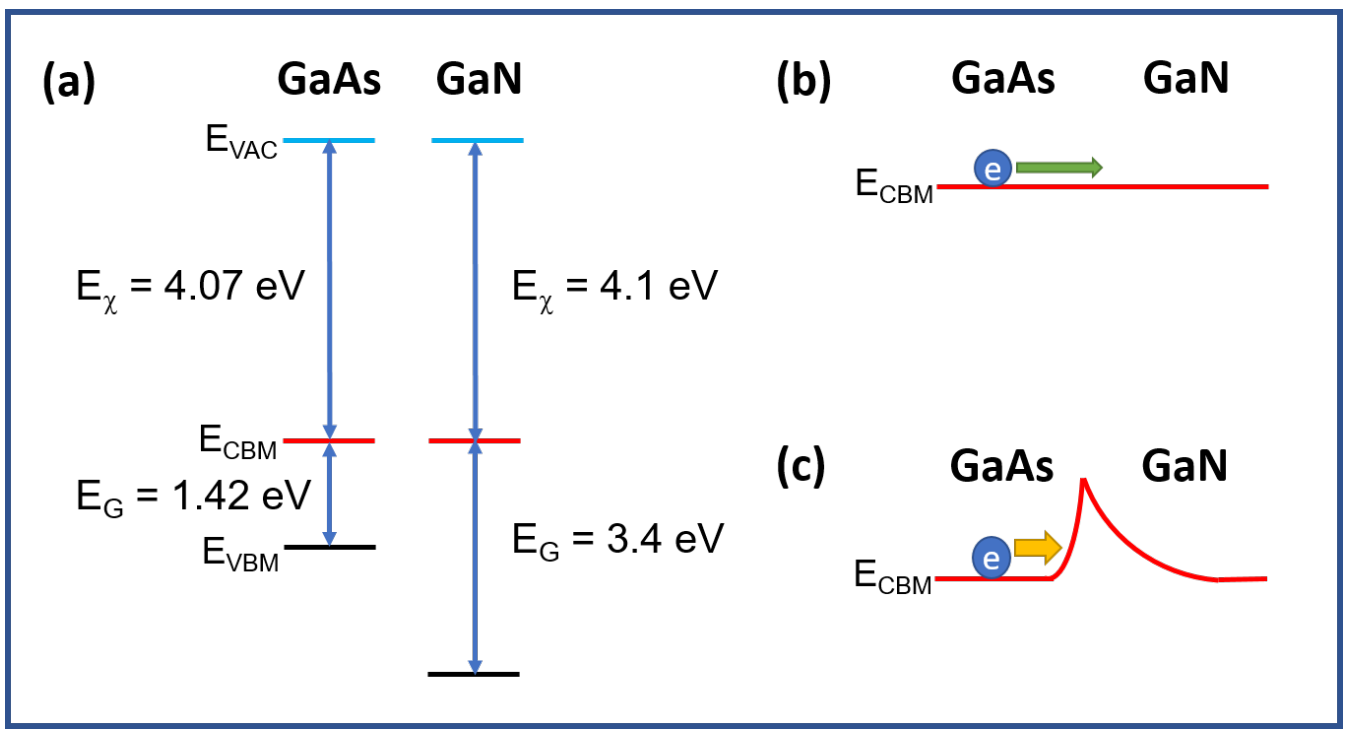

**Fig. 1**. Illustration of band alignment between GaAs and GaN and the influence of surface band-bending on electron transport from GaAs to GaN. (a) Band structures of GaAs and GaN (assuming no band-bending) before joining them together. Conduction band alignment between GaAs and GaN (b) without polarization induced band-bending and (c) with polarization induced band-bending in GaN. Due to the band-bending (c), electron transport from GaAs to GaN is blocked.

While the UO Al$_2$O$_3$ has been shown to play a critical role in the grafting method to achieve epitaxy-like lattice-mismatched heterojunctions between good p-type semiconductors and n-type GaN, achieving GaN-collector heterojunction bipolar transistors (HBTs) via grafting is not straightforward unless proper band alignment is realized between the constituent materials. For a npn GaN-collector HBT, like an AlGaAs/GaAs/GaN HBT, the conduction band alignment between the three materials needs to facilitate electron transport from emitter to collector. Since GaN exhibits surface energy band-bending due to polarizations and other associated effects., the band alignment of a grafted AlGaAs/GaAs/GaN may not favor the electron transport for HBT action. Fig. 1(a) shows the band diagrams of GaAs and GaN before they were grafted. Without considering the band-bending of GaN, GaN has an electron affinity of 4.1 eV [16] and a bandgap energy of 3.4 eV [17]. By grafting GaAs with GaN to form an epitaxial quality interface, the expected band alignment between the GaAs and GaN is illustrated in Fig. 1(b) without considering polarization. However, due to the polarization of GaN (c-plane, Ga-face) and the consequent upward band-bending, an energy barrier was formed at the conduction band between GaAs and GaN, as shown in Fig. 1(c). The energy barrier height is predominantly related to the GaN polarization strength, which can be affected by many factors. As a result, if forming/fabricating a npn AlGaAs/GaAs/GaN HBT via grafting without considering the band-bending of GaN, the energy barrier at the base-collector junction would block the electron current transport from the base region (p-GaAs) to the collector region (n-GaN), leading to poor or no transistor actions.

Previous studies have shown that the Al$_2$O$_3$/Ga-face GaN interface in metal-oxide-semiconductor (MOS) capacitors has a net positive charge [18-20] while a 0.4 eV upward band-bending was reported for the UO Al$_2$O$_3$-coated Ga-face GaN surface [21]. Although the mechanism of charge compensation at the Al$_2$O$_3$/Ga-face GaN interface is still under debate [22], these studies indicated the possibility to modulate the fixed charge density on the surface by UO Al$_2$O$_3$ and thus to reduce the surface band-bending of the c-plane, Ga-face GaN, potentially making GaN more suitable for HBT application. However, the results of these previous studies are not applicable to the envisioned GaN-collector HBT applications that can be enabled by the grafting approach due to the different interface treatment conditions (e.g., oxide thickness, annealing temperature, etc.). Therefore, it is imperative to perform a thorough examination of the influence of UO $Al_2O_3$ and related thermal process, which is relevant to grafting for GaN-collector HBTs, on the band-bending of c-plane, Ga-face GaN.

In this work, we present the surface band-bending studies of c-plane, Ga-face GaN grown on a sapphire substrate with different UO $Al_2O_3$ thicknesses that is directly relevant to grafting, ranging from 0 to 20 cycles by atomic layer deposition (ALD) with/without post-deposition annealing (PDA). The PDA parameters were selected based on the optimized processing conditions for grafting [11]. X-ray photoelectron spectroscopy (XPS) was employed to measure the valence band maximum (VBM) energy for calculating band-bending and with which surface band-bending values were derived. To help understand the implication of the band-bending modification, preliminary simulations of an AlGaAs/GaAs/GaN npn HBT with the measured band-bending values from this study were performed using Silvaco®.

**Experiment**

The epitaxial n-GaN (~ 250 nm thick) employed in this study was epitaxially grown on a two-inch sapphire substrate with a doping concentration (Si dopants) of $1 \times 10^{18}$ cm$^{-3}$, which was characterized by secondary-ion mass spectroscopy (SIMS). With this doping concentration, the Fermi level ($E_F$) was calculated to be 0.005 eV (($E_{CBM} - E_F$)$_{GaN}$ or $q\phi_{B,GaN}$) below the conduction band minimum (CBM). In this study, the epitaxial n-GaN wafer was used throughout the study under seven different surface treatment recipes. The cleaning of the n-GaN epi for all recipes of the study follows identical procedures: acetone, IPA, and DI water for 10 minutes, followed by 5 minutes of piranha etching (96% $H_2SO_4$ : 30% $H_2O_2$ : $H_2O$ = 1:1:8), 10 minutes of diluted hydrochloric acid etching (0.1 normal HCl : $H_2O$ = 1:1), and

5 minutes of diluted hydrofluoric acid etching (49% HF : $H_2O$ = 1:1) to remove the surface oxide (all the ratios are based on volume). For recipe #1-start, the n-GaN epi, which was a bare sample case without UO $Al_2O_3$, was immediately loaded for XPS analysis after cleaning. For recipes #2 to #7, the epi sample was loaded into the ALD chamber after cleaning. For ALD depositions, the ALD stage was heated to 200 °C and 5 cycles of trimethylaluminum (TMA) precursor flow was run with the purpose of in-situ native oxide removal, followed by $Al_2O_3$ deposition using the TMA precursor and $H_2O$ vapor under the same temperature of 200 °C. Each recipe has a different number of ALD deposition cycles. The expected ALD $Al_2O_3$ growth per cycle (GPC) is ~ 0.1 nm/cycle, although not strictly accurate for the initial cycles [23]. For recipes #2, #4, #6 and #7, XPS analyses were conducted directly after ALD deposition. For recipes #3 and #5, the sample was annealed in $N_2$ ambient at 350 °C for 5 minutes (post-deposition annealing (PDA)) before performing XPS analyses. After finishing the XPS analysis of recipe #7 (20 ALD deposition cycles), the $Al_2O_3$ of the epi sample was removed to make it a bare sample (recipe #1-end) again, and the bare n-GaN was analyzed using XPS. All ALD deposition and PDA conditions are identical to those used in the grafting of GaAs/GaN [11]. Table 1 summarized the detailed conditions of each recipe. Atomic force microscopy (AFM) analyses were conducted on the surface of the epi sample after each surface treatment using recipes #1, #2 and #3 (Fig. S1), revealing unchanged surface roughness. Combined with the XPS spectra of Al 2p (Fig. S1), it is confirmed that a smooth layer of $Al_2O_3$ was deposited on top of the GaN, consistent with the high-resolution transmission electron microscope (HRTEM) images of the grafted pn diodes [11]. Raman spectroscopy analyses were performed to assess the strain status of the n-GaN on sapphire epi sample under surface treatment recipe #1 and #7 (Fig. S2). It was found that the same amount of compressive strain exists for both recipes, matching the reported value of GaN epitaxial materials grown on sapphire substrates [24].

The XPS characterizations were performed using a Thermo Scientific K Alpha X-ray Photoelectron Spectrometer with Al $K_\alpha$ X-ray source ($hv$ = 1486.6 eV). The following settings were applied to the spectrometer: 10 eV pass energy, 100 μm spot size, 40s dwell time, and 0.02 eV step size. The spectrometer was calibrated using the Au $4f_{7/2}$ peak and the charge correction was conducted using the C 1$s$ peak at 284.6 eV [25]. The Au $4f_{7/2}$ spectrum was used to calculate the instrumental uncertainty. By fixing the full width at half maximum (FWHM) Lorentzian lifetime width Γ within the range of 0.30 eV to 0.34 eV for the Au $4f_{7/2}$ spectrum, a total Gaussian resolution σ, which includes the uncertainty from Gaussian phonon and that from the spectrometer itself, of 0.19 eV to 0.21 eV was fitted [25-27].

**Results and discussion**

Fig. 2(a) shows the recorded XPS spectrum of the n-GaN valence band directly from the XPS spectrometer under surface treated using recipe #1-start. The spectrum is also representative of all XPS spectra acquired from the studied recipes (Fig. S3). Based on the XPS spectra obtained, the binding energy of VBM at the surface of GaN (($E_F - E_{VBM})_{GaN}$ or $q\phi_{S,GaN}$) was determined using the method developed by Reddy Pramod et al. [25, 28], as shown in Fig. 2(b). The theoretical density of state (DOS) for 3D hexagonal crystal lattice with consideration of nearby atoms can be calculated using the Monte Carlo method in the first Brillouin zone using the energy dispersion relation

$$E = \epsilon - t\left(2\cos(k_x a) + 4\cos(\sqrt{3}k_y a/2)\cos(k_x a/2) + 2\cos(k_z c)\right), \quad (1)$$

where $\epsilon$ is the reference energy, $t$ is the bond energy, and $a$ and $c$ are the lattice constants of GaN. This theoretical DOS (the red curve in Fig. 2(b)) is a good approximation of the XPS spectrum around the VBM (the black curve in Fig. 2(b)) after Gaussian broadening [25]. The DOS curve was then convoluted with a Gaussian broadening equation whose standard deviation is the total Gaussian resolution σ of the spectrometer. By matching the convoluted DOS curve (the blue curve in Fig. 2(b)) with the acquired XPS spectrum (the black curve in Fig. 2(b)), the bond energy $t$ was determined to be -0.2 eV. $q\phi_{S,GaN}$ can be read directly from the lower edge on the theoretical DOS curve (Fig. 2(b)) after obtaining a good fit between the convoluted DOS curve and the measured XPS spectrum around the VBM. For recipe #1-start, $q\phi_{S,GaN}$ is 2.91 eV (Fig. 2(b)). Combining the GaN bandgap energy ($E_{G,GaN}$) of 3.4 eV and the calculated $q\phi_{B,GaN}$ of 0.005 eV, an upward band-bending ($q\psi_S$) of 0.48 eV was obtained for recipe #1-start from the following equation.

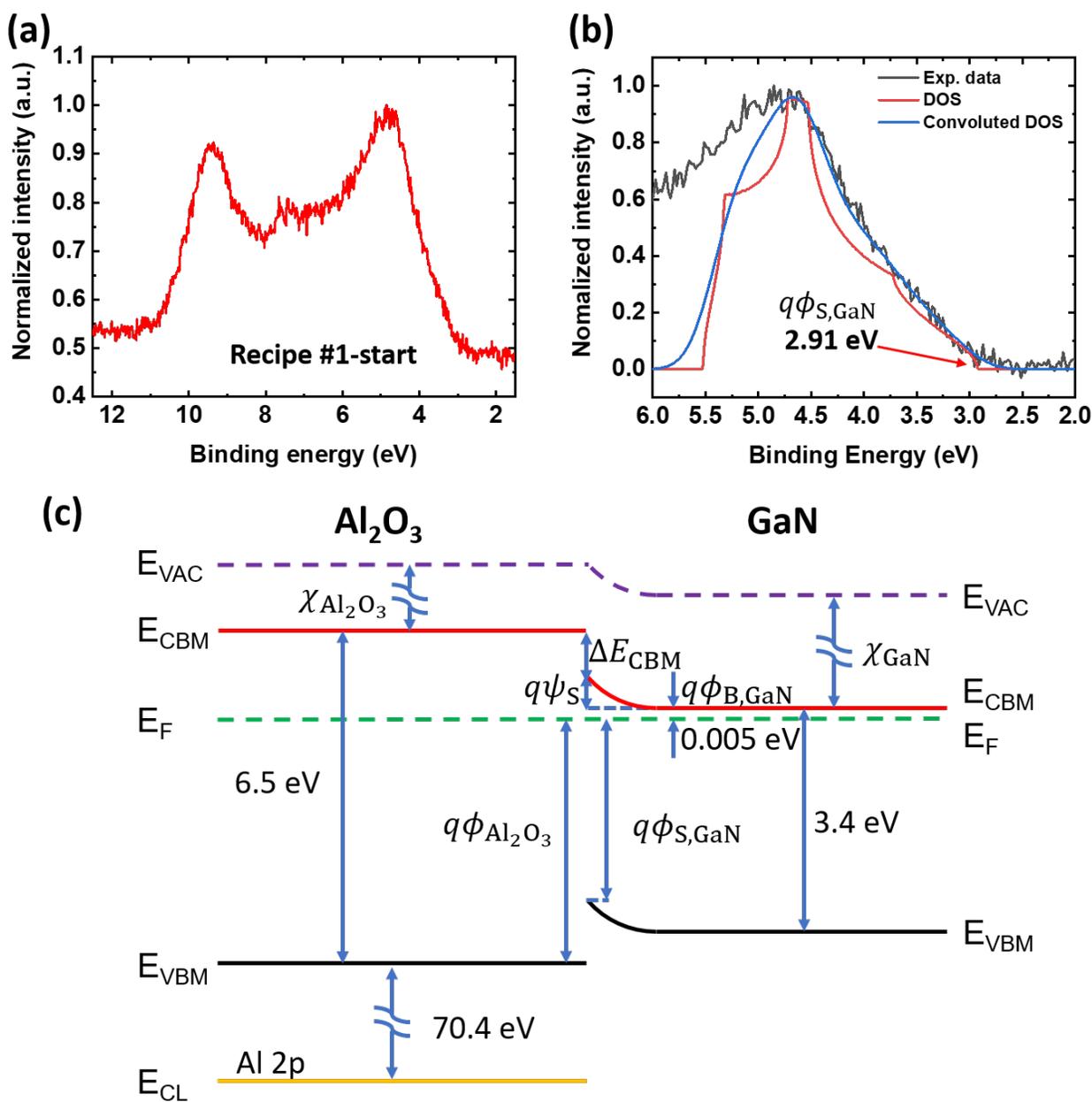

**Fig. 2**. (a) The valence band XPS spectrum of n-GaN epi (recipe #1-start) directly recorded from the spectrometer. (b) XPS spectrum with a Shirley background subtracted (black line), theoretical DOS curve (red line), and theoretical DOS curve convoluted with the instrumental Gaussian broadening equation (blue line) around the VBM. (c) A Schematic illustration of the band alignment between $Al_2O_3$ and GaN.

**Table 1**. Summary of the sample treatment conditions, measured $q\phi_{S,GaN}$, upward band-bending energy values $q\psi_S$, and conduction band barrier height $\Delta E_{CBM}$.

| Recipe # | Native oxide removal | ALD cycles (CY) | PDA | $q\phi_{S,GaN}$ (eV) | $q\psi_S$ (eV) | $\Delta E_{CBM}$ (eV) |
|---|---|---|---|---|---|---|
| 1 (Start) | Yes | - | - | 2.91 | +0.48 | - |
| 2 | Yes | 5 | - | 2.96 | +0.44 | 1.57 |
| 3 | Yes | 5 | 350°C 5 min | 3.03 | +0.36 | 1.67 |
| 4 | Yes | 10 | - | 3.03 | +0.37 | 1.85 |
| 5 | Yes | 10 | 350°C 5 min | 3.09 | +0.30 | 1.75 |
| 6 | Yes | 15 | - | 3.13 | +0.27 | 1.86 |
| 7 | Yes | 20 | - | 3.28 | +0.12 | 2.13 |
| 1 (End) | Yes | - | - | 3.00 | +0.39 | - |

$$q\psi_S = E_{G,GaN} - q\phi_{S,GaN} - q\phi_{B,GaN}, \quad (2)$$

In this article, we define upward band-bending as positive. Using the identical methods, the $q\phi_{S,GaN}$ and the $q\psi_S$ values of the n-GaN epi under treatment of other recipes were obtained. The $q\psi_S$ values are listed in Table 1 and are also plotted in Fig. 3(a) as a function of the ALD Al$_2$O$_3$ deposition cycles.

From the measured XPS spectra, the barrier height $\Delta E_{CBM}$ of Al$_2$O$_3$ can be calculated using the following relation according to the illustration in Fig. 2(c),

$$\Delta E_{\text{CBM}} = \left(E_{\text{G},\text{Al}_2\text{O}_3} - q\phi_{\text{Al}_2\text{O}_3}\right) - q\psi_{\text{S}} - (E_{\text{CBM}} - E_{\text{F}})_{\text{GaN}}$$

$$= \left(E_{\text{G},\text{Al}_2\text{O}_3} - [E_{\text{CL}} - (E_{\text{CL}} - E_{\text{VBM}})]_{\text{Al}_2\text{O}_3}\right) - q\psi_{\text{S}} - q\phi_{\text{B},\text{GaN}}, \quad (3)$$

where $\Delta E_{\text{CBM}}$ is the conduction band barrier between $Al_2O_3$ and GaN, $E_{\text{G}}$ is the bandgap, $q\phi_{\text{Al}_2\text{O}_3}$ is the energy difference between Fermi level and VBM in $Al_2O_3$, $E_{\text{CL}}$ is the binding energy (referred to $E_{\text{F}}$) of the core level (e.g., Al 2p), and $E_{\text{CL}} - E_{\text{VBM}}$ is the energy difference between Al 2p core level and the amorphous ALD $Al_2O_3$ VBM. As mentioned above, $E_{\text{CL},\text{Al}_2\text{O}_3}$ is selected as Al 2p level, and it has been measured from XPS (Fig. S1). For amorphous ALD $Al_2O_3$, $E_{\text{G}}$ is adopted as 6.5 eV [21, 29-31], and $E_{\text{CL}} - E_{\text{VBM}}$ is 70.4 eV [21]. The conduction band barrier height value varies for each recipe and can be calculated, respectively (Table 1). And the band barrier from $Al_2O_3$ can affect the electrical characteristics of junctions containing interface $Al_2O_3$, as shown previously [11].

It is worth noting that in the work by Reddy Pramod et al. [25], the bond energy $t$ is chosen to be positive, corresponding to an energy dispersion relation having Γ point as the minimum energy point. Therefore, the energy dispersion relation with a positive *t* describes the CBM in theory, but not applicable for VBM. While using a positive bond energy $t$ can theoretically lead to incorrect band-bending values, it may not be obvious when a low energy resolution of the instrument (e.g., ~ 0.5 eV in ref. 24) was used, which can smooth out most of the asymmetric features of the theoretical DOS curve after convolution with the Gaussian broadening equation.

As shown in Fig. 3(a), a monotonic trend of the reduction of the upward surface band-bending energy from 0.48 eV down to 0.12 eV was observed as the number of UO $Al_2O_3$ deposition cycles increased from 0 to 20 cycles. Every additional five-cycle ALD $Al_2O_3$ deposition resulted in additional band-bending reduction, ranging from 70 meV (from 10 cycles to 5 cycles), 100 meV (from 15 cycles to 10 cycles), and 150 meV (from 20 cycles to 15 cycles). Each PDA procedure caused further band-bending reduction, ranging from 70 meV (5 cycles annealed) to 60 meV (10 cycles annealed). We expect that the same amount of band-bending reduction may be observed for other cases under the same PDA condition [32].

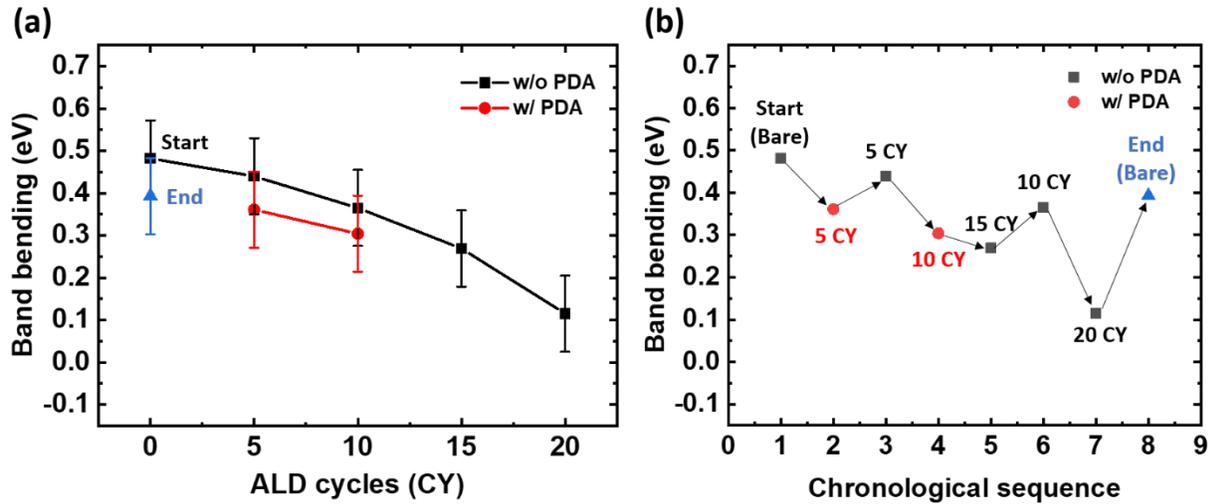

**Fig. 3**. (a) Plot of the measured surface band-bending values of the n-GaN epi as a function of the number of ALD Al$_2$O$_3$ cycles. The lines between the data points are drawn to guide the view of the data. The error bars were estimated based on the standard deviation of Al 2p XPS spectra (0.09 eV). (b) Replot of the data points shown in (a) as a function of the chronological sequence index of the experiment. The arrows between every two data points indicate the experiment sequence.

Since all the surface treatments were performed on the same n-GaN epi sample, to examine how the surface treatments alter the surface conditions of the sample after a series of treatments, the band-bending energy values were replotted as a function of the chronological sequence order, as shown in Fig. 3(b). It is noted that the chronological sequence was not designed intentionally to follow a rough increasing order of the increased number of ALD deposition cycles, instead; it was summarized after all experiments were finished. Fig. 3(b) shows no clear trend of any time-dependence other than the trend that was already shown in Fig. 3(a). Of most importance and interest is the surface band-bending's approximate "recovery" (0.39 eV for the bare/recipe #1-end case) of the n-GaN epi after removing the 20-cycle UO Al$_2$O$_3$ layer that was deposited in recipe #7. The band-bending "recovery" indicates that most of the observed modulation of the upward band-bending (5-20 cycles/recipe #2-#7) was the effect caused by the deposited UO $Al_2O_3$. The remaining difference (0.09 eV) between recipe #1-start and recipe #1-end could be due to the uncertainty of the experiments or, to a small speculated extent, due to the removal/oxidation of some Ga atoms from the epi surface and thus the associated change of the density of polarization charges at the surface caused by the series of surface treatments. Nevertheless, Fig. 3 clearly indicates that UO $Al_2O_3$ deposition on GaN surface can reduce the upward band-bending of Ga-face, c-plane GaN, and more cycles of $Al_2O_3$ deposition can have more effect on the reduction of the upward band-bending.

There are multiple theories about charge compensation by ALD $Al_2O_3$ on III-nitride surface. The community has reported in metal-oxide-semiconductor (MOS) GaN structure that the interface fixed charge between $Al_2O_3$ and III-nitride can be changed from intrinsic negative charge to positive charge [18-20, 32, 33] and proposed multiple hypotheses of the mechanisms. We analyzed these reported theories to identify the most plausible one that is applicable to our results. Theory #1: Esposto et al. proposed the charge compensation theory based on the polarization inversion due to the Ga-O bonds formation [18]. However, Ga-O bonds exist on the measured surfaces from recipe #2 to #7 (see Fig. S4) and thus they are unlikely to be the reason for the change of surface fixed charges in this work. Furthermore, polarization inversion was not observed in this study. Theory #2: Ganguly et al. and Ťapajna et al. treated oxygen atoms as donor dopants in GaN, which contribute positive fixed charges at the surface of GaN [32, 33]. But for recipes #2 to #7, all surfaces have more than the sufficient number of oxygen atoms to substitute the nitrogen sites on the surface [33], meaning that the same band-bending should be measured through these recipes if the theory holds true, which is also contrary to the results observed in this work. Theory #3: Ťapajna et al. and Hayashi et al. developed the hypothesis that donor-type surface defects act as positive fixed surface charges on GaN [19, 34]. However, the trap state density on the surface with ALD $Al_2O_3$ has been experimentally found to decrease with the increasing number of ALD cycles previously

[35, 36]. Therefore, one would expect to see a weaker compensation effect with more cycles of ALD $Al_2O_3$. This is opposite to what we have observed in our work. Theory #4: Bakeroot et al. proposed the border trap theory [37]. According to this theory, if we assume a uniform density of traps inside the insulator ($Al_2O_3$) along the deposition direction, about 5 nm of $Al_2O_3$ would be needed to achieve the polarization inversion that has been observed in a GaN MOS structure, based on the trend observed in this work. As a matter of fact, it is questionable whether the insulator traps located at this distance away from the interface can still have influence on GaN. Theory #5: Esposto et al. and Matys et al. proposed that the charge compensation mechanism at the energy states between the CBMs of the UO and that of GaN [18, 38] can cause the change of band-bending. The band structure and the density of states in the band gap of UO $Al_2O_3$ can possibly change with different numbers of cycles therefore, providing different degrees of charge compensation effects. It is therefore believed that the charge compensation mechanism as described in the theory by Esposto et al. [18] and Matys et al. [38] is the most probable one that is responsible for the observed band-bending in this work.

The demonstrated modulation/suppression of energy band-bending at the n-GaN surface enabled by the UO $Al_2O_3$ implies the potential to develop practical AlGaAs/GaAs/GaN HBTs employing the grafting approach [11]. To illustrate this implication, an AlGaAs/GaAs/GaN npn HBT with a vertical layer structure shown in Fig. S5(a) and Fig. S6(a) was simulated using Silvaco® device simulator Atlas® (Fig. S5(b) and Fig. S6(b)). The emitter and base layers of the HBT used in the simulations are identical to those of the AlGaAs/GaAs/GaN HBT fabricated using wafer fusion [8]. The only major difference between the fused HBT [8] and the simulated HBT is the UO $Al_2O_3$ interface between the p+ GaAs (base) and the n- GaN (collector) in the latter case. To perform the HBT simulations, the UO $Al_2O_3$ at the interface was treated in two ways. In Case 1 (Fig. S5), it is considered an ideal interface (i.e., ignoring the existence of the $Al_2O_3$) given the ideal p-n junction that can be achieved in grafting ($n \sim 1.09$ or less) [11]. In Case 2 (Fig. S6), the UO $Al_2O_3$ is considered an insulator. The detailed settings and descriptions of the simulations can be found in the Supplementary Information (SI). It is noted that the UO Al$_2$O$_3$ functions as a double-sided passivation and quantum tunneling layer. Therefore, neither of the two treatment methods accurately captures the physical charge carrier transport mechanisms in a real grafted GaAs/GaN pn junction at this stage of the simulations study.

Fig. S5(d) shows the simulated current gain ($\beta$) values of the HBT as a function of the base-emitter voltage ($V_{BE}$) at collector-emitter voltage ($V_{CE}$) of 5 V, under different surface treatment conditions, for Case 1. The observed increasing trend of the $\beta$ values as the surface band-bending energy of GaN, the only variable in the simulations, decreases reflects the direct correlation between GaN surface band-bending and the HBT transistor action. Since the effects of the UO Al$_2$O$_3$ layer were not taken into account in the simulations, the $\beta$ values obtained from the simulations are considered an overestimate.

Fig. S6(d) shows the simulated $\beta$ values of the HBT as a function of $V_{BE}$ under collector-base voltage $V_{CB}$ of 3 V for Case 2. It is noted that because the UO Al$_2$O$_3$ is considered as an insulator in this case, the polarization of the GaN cannot implemented in the simulator. The $\beta$ at higher $V_{CB}$ cannot be simulated due to convergence problems encountered in the simulations, which is further due to the insulator setting for the Al$_2$O$_3$ layer. In this case, only recipe #3 can lead to a useful $\beta$ value (30). Given that the UO Al$_2$O$_3$ layer is a conductor layer, treating the layer as an insulator has underestimated the $\beta$ values of the HBT on one hand. On the other hand, the setting of no polarization should have overestimated the $\beta$ value. While a more accurate modeling approach that can better depict the physical behavior of the Al$_2$O$_3$ layer in the grafted system needs to be developed, the simulations under the two extreme conditions (Case I and Case II) together with the band-bending characterizations in this work have allowed one to peek into the potential of implementing AlGaAs/GaAs/GaN npn HBTs using the grafting approach.

It is noted that although AlGaAs/GaAs/GaN npn HBT was envisioned and used as the GaN-collector HBT example in the simulation study, the principle of the GaN-collector HBT work mechanism is also applicable to several other suitable material combinations (e.g., npn Si/Ge/GaN, npn GaAs/Si/GaN, npn GaAs/Ge/GaN etc., to name a few) provided that the conduction band alignment of the constituent materials is appropriated for electron transport in the bipolar transistor action mode and that, from a practical point of view, the material combinations are implementable via the grafting approach [11]. Future device fabrication work using the grafting approach will be needed to experimentally verify the GaN-collector HBT prospective, along with characterization of the interface such as interface trap distribution in the bandgap [39] to help us better understand the grafted interface with UO.

**Conclusion**

In summary, it is shown that the essential UO $Al_2O_3$ needed at the GaAs/GaN interface in grafting can help suppress the upward band-bending of c-plane, Ga-face GaN with a monotonic trend of suppression enhancement following the increase of the $Al_2O_3$ thickness. Employing 20 cycles ALD $Al_2O_3$ can suppress the band-bending of GaN from 0.48 eV of a bare GaN down to 0.12 eV. Preliminary device simulations show that the suppression of upward band-bending with proper UO thickness can potentially improve the current gain of AlGaAs/GaAs/GaN npn HBT with a grafted base-collector junction, implying the prospective of development of practical GaN-collector HBTs. New experimental work is in progress to further examine the electrical effects from both the band barrier of $Al_2O_3$ and the observed band-bending of GaN.

**Acknowledgement**

The work was supported by Air Force Office of Scientific Research under grant FA9550-21-1-0081.

Supplementary Information:

# Influences of ALD Al$_2$O$_3$ on the surface band-bending of c-plane, Ga-face GaN and the implication to GaN-collector npn heterojunction bipolar transistors

Jiarui Gong[1], Jisoo Kim[2], TienKhee Ng[3], Kuangye Lu[4], Donghyeok Kim[2], Jie Zhou[2], Dong Liu[2], Jeehwan Kim[4], Boon S. Ooi[3], and Zhenqiang Ma[2,a]

[1]*Department of Physics, University of Wisconsin-Madison, Madison, Wisconsin, 53706, USA*

[2]*Department of Electrical and Computer Engineering, University of Wisconsin-Madison, Madison, Wisconsin, 53706, USA*

[3]*Department of Electrical and Computer Engineering, King Abdullah University of Science and Technology, Thuwal 23955-6900, Saudi Arabia*

[4]*Department of Mechanical Engineering, Department of Materials Science and Engineering, Massachusetts Institute of Technology, Cambridge, Massachusetts, 02139, USA*

[a]    Author to whom correspondence should be addressed. Electronic mail: mazq@engr.wisc.edu

# I. Atomic Force Spectroscopy (AFM) images of the n-GaN Epitaxial Samples

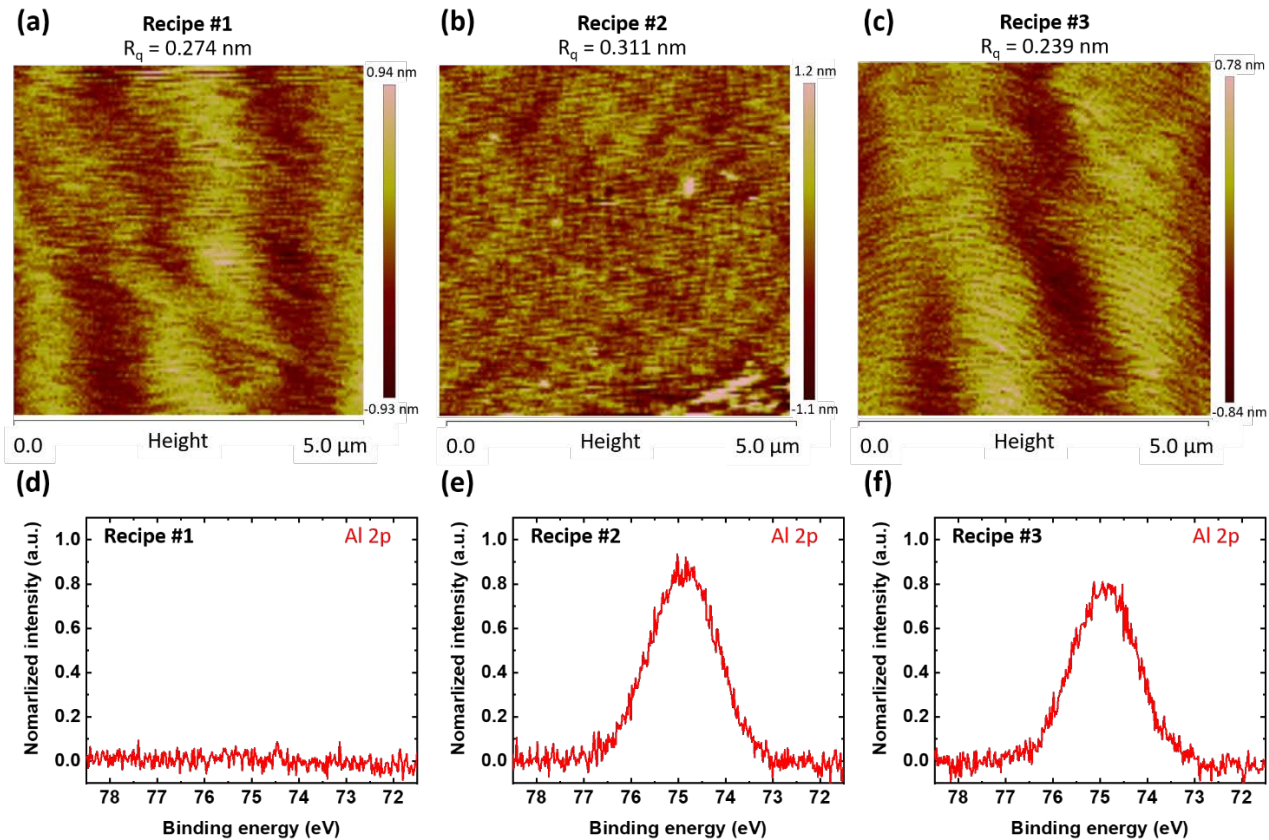

**Fig. S1**. AFM images and Al 2p XPS spectra of the n-GaN epi sample under different surface treatment conditions: (a) $R_q$ = 0.274 nm and (d) for recipe #1: bare surface, (b) $R_q$ = 0.311 nm and (e) for recipe #2: 5-cycle ALD $Al_2O_3$ coating, and (c) $R_q$ = 0.239 nm and (f) for recipe #3: 5-cycle ALD $Al_2O_3$ coating and PDA.

## II. Raman Spectroscopy Analysis of the n-GaN Epitaxial Samples

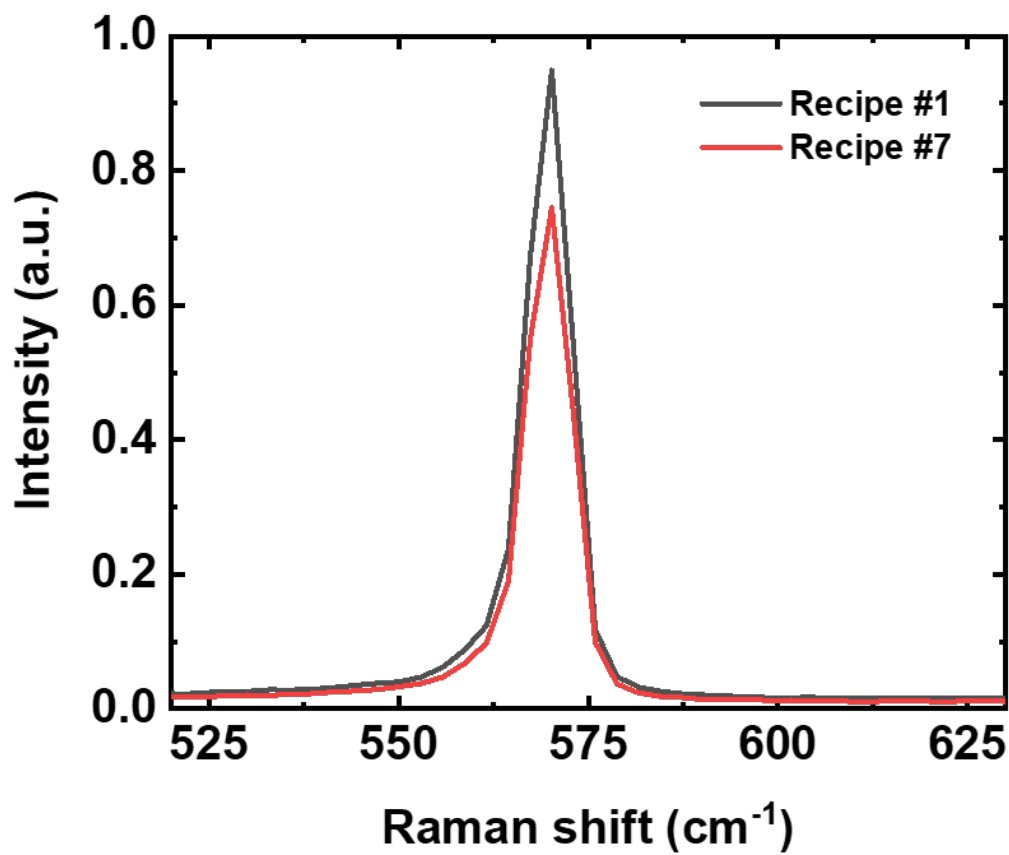

**Fig. S2**. Raman spectra of recipe #1 (bare) and recipe #7 (20 cycles).

We performed Raman spectroscopy analysis to assess the stress status of the n-GaN epitaxial layer. Raman spectra were performed using a Raman microscopic system (α 300M+, WITec) with a pump laser wavelength of 532 nm and an objective lens of 100× magnification. As shown in Fig. S2, we measured the frequency shift of the $E_2$ high ($E_2^H$) mode of GaN, which is a good representation of biaxial strain in c-plane GaN [1]. For relaxed GaN, the $E_2^H$ mode phonon frequency is observed at 568 cm$^{-1}$ [2]. For our n-GaN epi on sapphire, it is measured at 570 cm$^{-1}$ for both recipes #1 and #7, which matches the reported value for GaN on sapphire [1] and exhibits compressive stress. The results show that the ALD $Al_2O_3$ does not introduce additional stress to the GaN epi.

# III. XPS Spectra and fitting of the n-GaN Epitaxial Samples

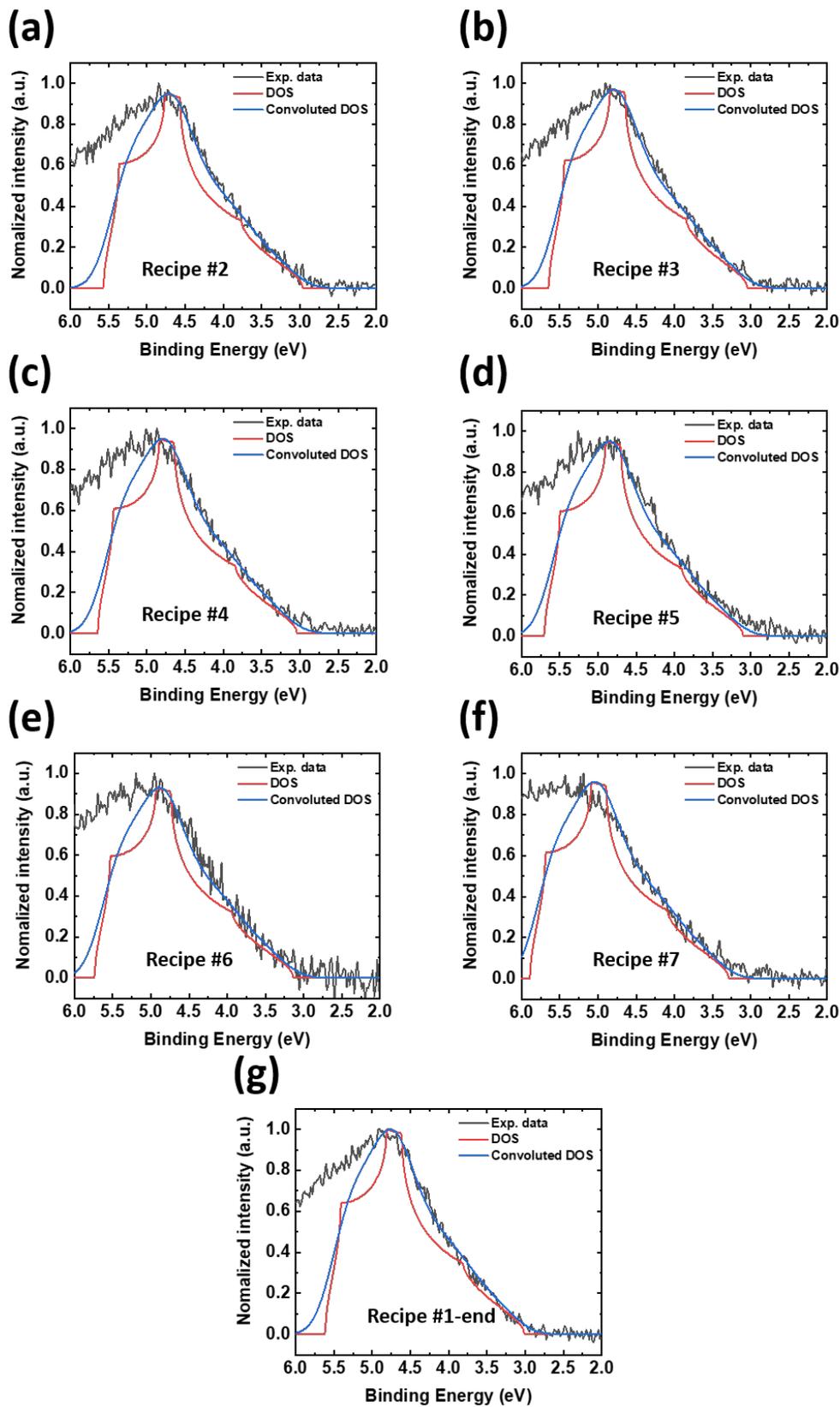

**Fig. S3.** XPS spectra with a Shirley background subtracted (black line), theoretical DOS curve (red line), and theoretical DOS curve convoluted with the instrumental Gaussian broadening equation (blue line) around the VBM for recipes #2 (a) to #7 (f) and recipe #1-end (g).

## IV. Ga 3d XPS Spectra of the n-GaN Epitaxial Samples

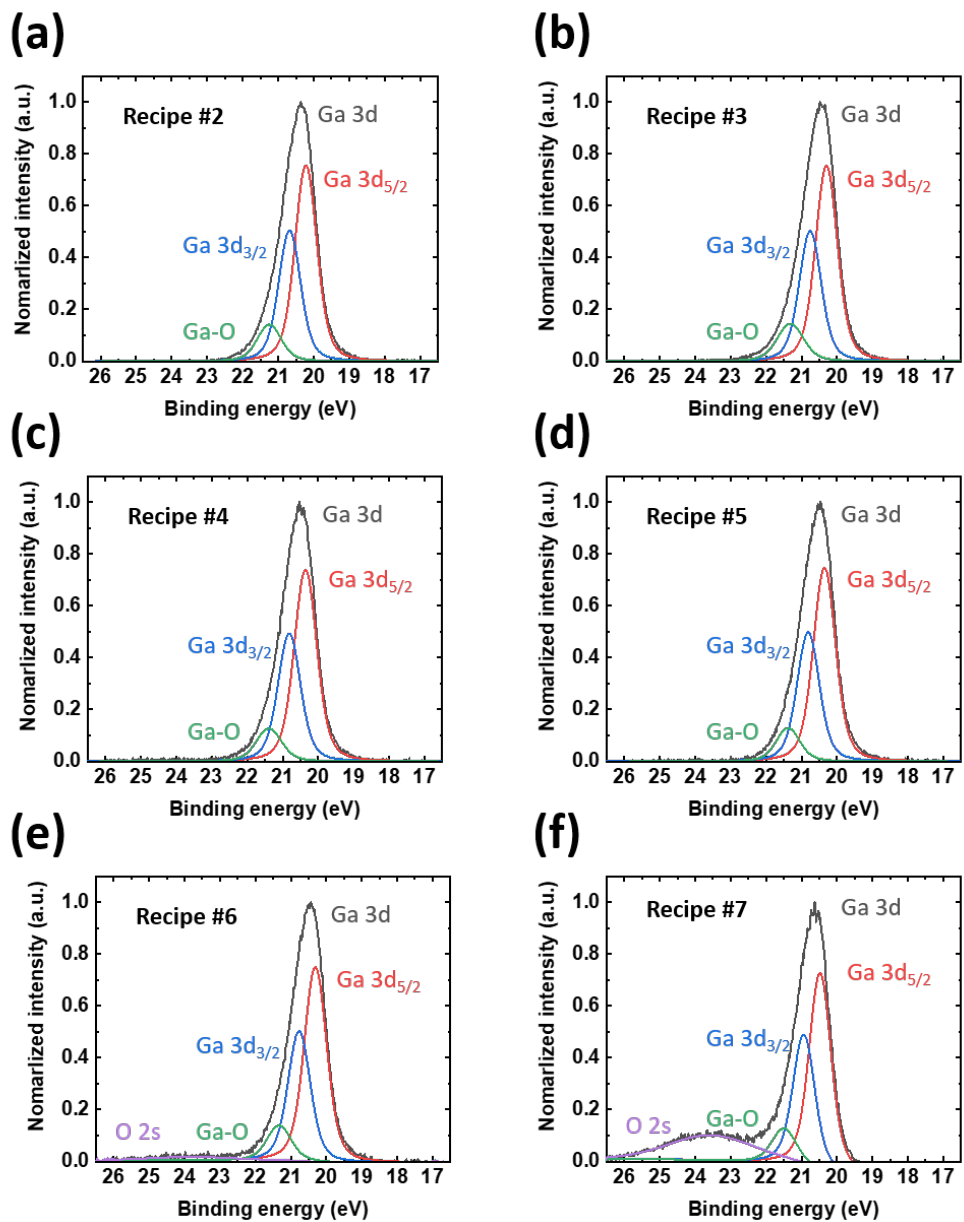

**Fig. S4.** Ga 3d XPS spectra of recipes #2 (a) to #7(f). A Sherley background is subtracted from each spectrum.

## V. AlGaAs/GaAs/GaN npn HBT Simulations

The present device simulators are unable to accurately model the charge carrier transport across the ultrathin dielectric layer ($Al_2O_3$) used in grafted semiconductor material systems because the interface $Al_2O_3$ layer is treated as an insulator, while the roles of the $Al_2O_3$ layer are double-side passivation and quantum tunneling glue layer allowing both electrons and holes to tunnel through [3]. As a result, the $Al_2O_3$ layer behaves like a super-wideband semiconductor, forming double interfaces with its neighboring semiconductors with finite (very low) densities of states. Based on these considerations, we provide simulation results of the AlGaAs/GaAs/GaN npn HBT under two extreme conditions using Silvaco® device simulator Atlas®: 1) treating the GaAs/GaN interface as an ideal interface (i.e., ignoring the existence of the $Al_2O_3$ layer), under which condition the HBT performance should be overestimated; and 2) treating the $Al_2O_3$ layer as an insulator, under which condition the HBT performance should be underestimated due to the poorer conductivity of the insulator than the reality, where the $Al_2O_3$ layer is an effective electron/hole conductor. We expect that an accurately predicted HBT performance, upon the development and availability of a suitable model for modeling the grafted material system, would lie in between the results of these two extreme conditions.

*Case I: Treating the GaAs/GaN interface as an ideal interface*

Since the grafted interface between GaAs and GaN can have an ideal p-n junction ($n \sim 1.09$ or less) [3], the GaAs/GaN junction was treated as an epitaxy-like junction in the simulations in this case. Under this ideal assumption, the only material variable parameter in the simulations is the polarization scaler of the GaN, which was set according to the measured corresponding band-bending values listed in Table 1 (also shown in Fig. 3).

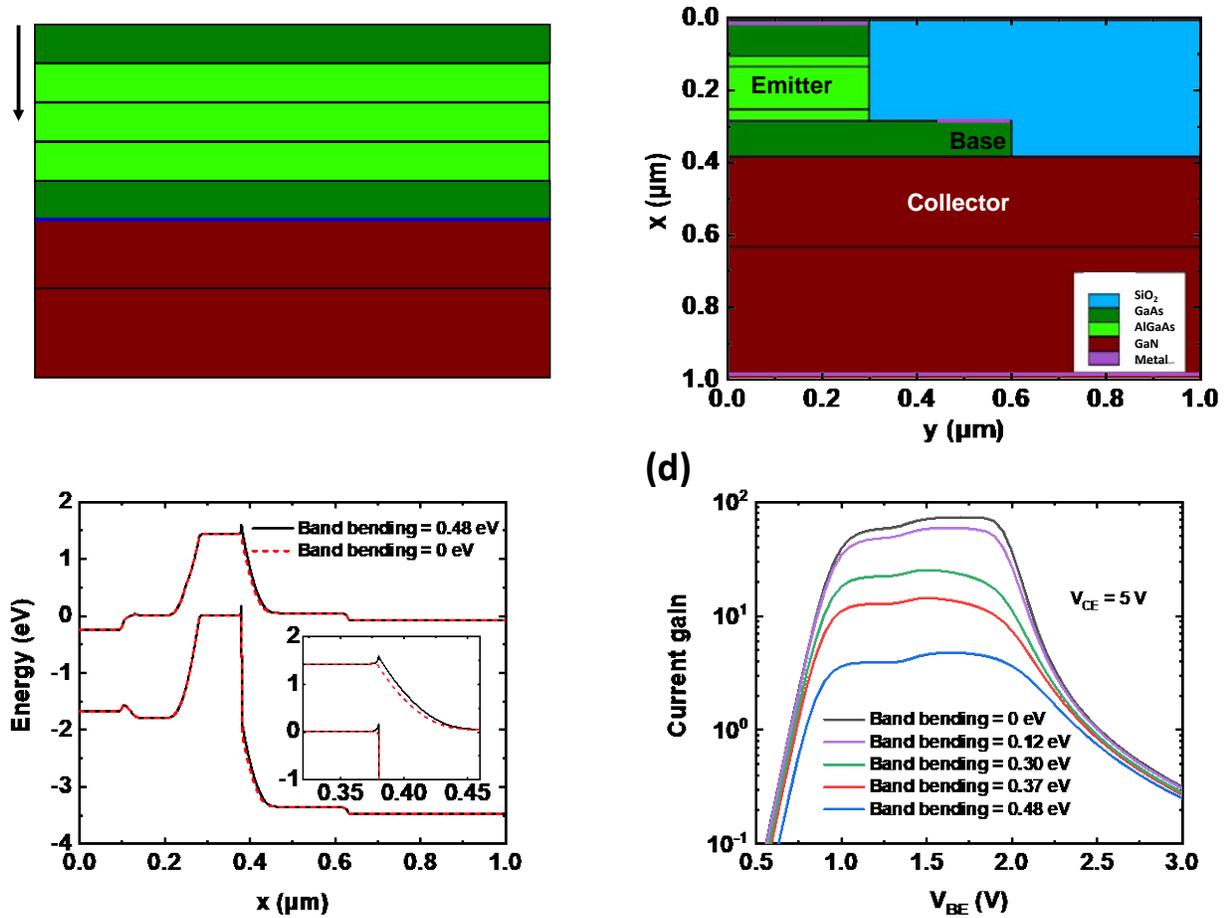

**Fig. S5**. Case I. (a) Schematic layer structure of an AlGaAs/GaAs/GaN npn HBT with ALD Al$_2$O$_3$ at the interface of GaAs/GaN. (b) Simulated AlGaAs/GaAs/GaN npn HBT structure at the interface of Silvaco®. (c) Simulated band diagram at zero bias of the AlGaAs/GaAs/GaN npn HBT with 0.48 eV band-bending (black solid lines) and 0 eV band-bending (red dashed lines). The inset is the enlarged region around the GaAs/GaN heterojunction showing the detailed band structure at the GaAs/GaN interface. (d) Simulated current gain $\beta$ versus base-emitter voltage $V_{BE}$ at $V_{CE}$ of 5 V under different n-GaN upward band-bending values. In the simulations, only the band-bending energy value varied.

Fig. S5(a) shows a schematic illustration of the simulated HBT structure (the Al$_2$O$_3$ layer was only treated as an ideal double-side passivation layer and not inserted physically in the simulation). The emitter and base layers of the HBT (Fig. S5(a)) used in the simulations are identical to that of the

AlGaAs/GaAs/GaN HBT fabricated using wafer fusion [4]. Fig. S5(b) shows the HBT structure in Atlas® interface. Fig. S5(c) shows the simulated band diagrams of the HBT with the n-GaN as the HBT collector having an upward band-bending of 0.48 eV (recipe #1-start) and having zero upward band-bending (an imagined extreme situation) for comparison. The inset of Fig. S5(c) shows an enlarged view of the comparison of the band alignment at the base-collector junction under the two band-bending situations. The simulated current gain ($\beta$) values of the HBT as a function of the base-emitter voltage ($V_{BE}$) at collector-emitter voltage ($V_{CE}$) of 5 V, under different upward band-bending energies, are plotted in Fig. S5(d). From Fig. S5(d), one can see that the GaN polarization-induced upward band-bending substantially reduces the HBT current gain from 74 (0 eV band-bending) to 4.8 for the bare GaN (0.48 eV band-bending). The results indicate that with an ideally passivated interface between GaAs and c-plane, Ga-face GaN, like what was already experimentally achieved [3], developing practical GaN-collector HBT is nearly impossible due to the electron transport barrier created by upward band-bending of the c-plane, Ga-face GaN. Nevertheless, the UO $Al_2O_3$ at the interface of GaAs and GaN with proper thicknesses could lead to both an ideal pn junction and a suitable band alignment at the p-GaAs/n-GaN base-collector junction, facilitating electron transport from the AlGaAs emitter, crossing the GaAs base, and collected by the GaN collector. Specifically, with a 20-cycle ALD $Al_2O_3$ layer (i.e., 0.12 eV upward band-bending as measured from the XPS in this study), the $\beta$ can reach 59, slightly lower than the ideal case (no band-bending) but 12 times higher than that of bare GaN. Although the effects of the UO $Al_2O_3$ on the electrical characteristic of the HBT have not been considered, the simulations together with the band-bending characterizations in this work have shed light on the promising route toward GaN-collector HBTs.

*Case II. Treating the $Al_2O_3$ layer as an insulator*

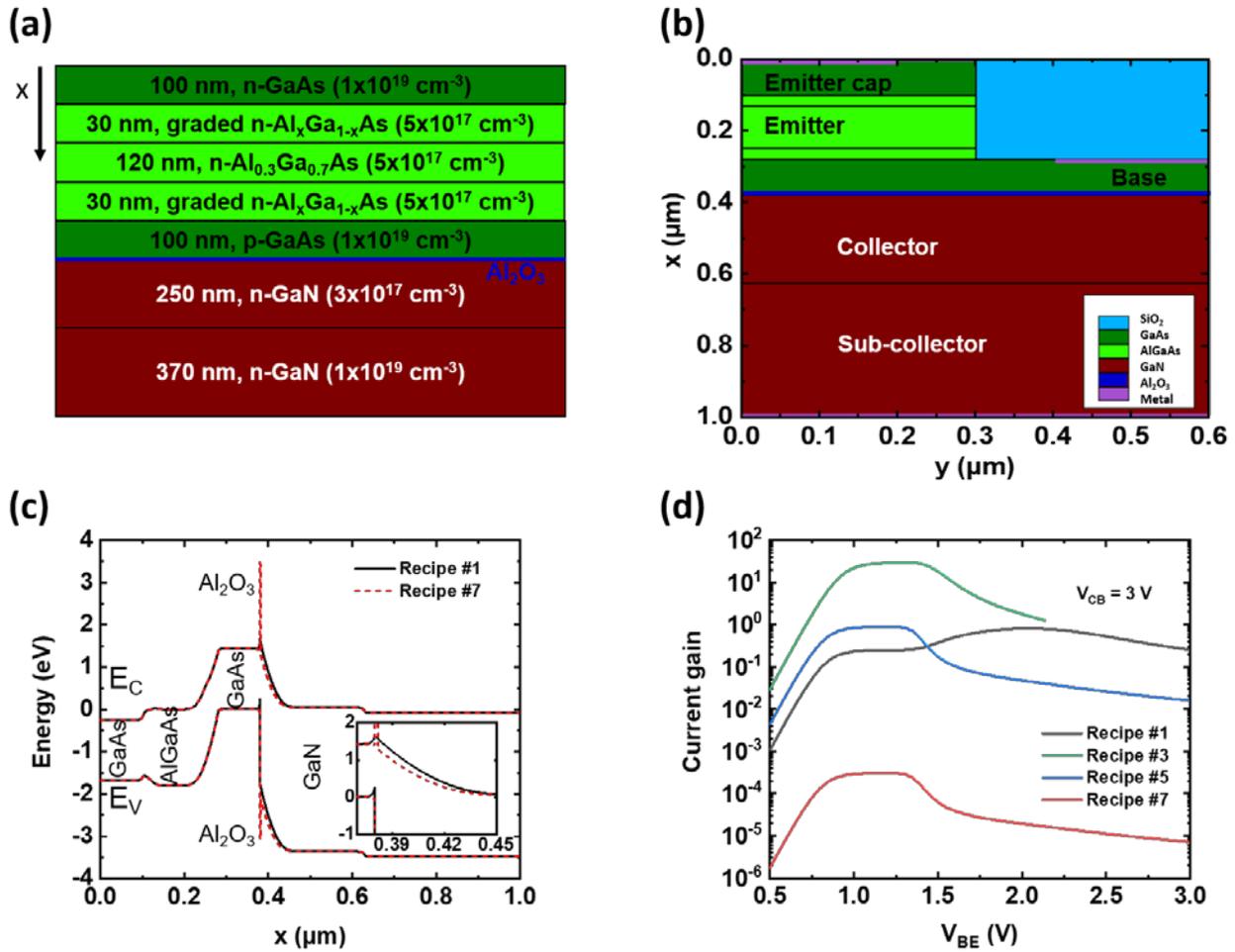

**Fig. S6**. Case II. (a) Schematic layer structure of an AlGaAs/GaAs/GaN npn HBT with ALD $Al_2O_3$ (as insulator) at the interface of GaAs/GaN. (b) Simulated device structure at the interface of Silvaco®. (c) Simulated band diagram at zero bias of the AlGaAs/GaAs/GaN npn HBT of surface treatment recipe #1 (black solid lines) and recipe #7 (red dashed lines). The inset is the enlarged region around the GaAs/GaN heterojunction showing the detailed band structure at the GaAs/GaN interface. (d) Simulated current gain $\beta$ versus base-emitter voltage $V_{BE}$ at $V_{CB}$ of 3 V under different surface treatments.

In this case, the identical layer structure to Fig. S5(a) was used, except that the $Al_2O_3$ interlayer was inserted as an insulator layer and the tunneling models (sis.el and sis.ho) were applied to the base/collector junction in the simulation (Fig. S5(b)).

Fig. S6(c) shows the simulated band diagram of the HBT with n-GaN as the HBT collector having an upward band-bending of 0.48 eV (recipe #1-start) and 0.12 eV (i.e., the 20-cycle $Al_2O_3$ coated GaN collector: recipe #7)) for comparison. The inset of Fig. S6(c) shows an enlarged view of the comparison of the band alignment at the base-collector junction. The simulated $\beta$ values of the HBT as a function of $V_{BE}$ at collector-base voltage ($V_{CB}$) of 3 V, under different surface treatment conditions, are plotted in Fig. S6(d). From Fig. S6(d), one can see that a balance between effective quantum tunneling across the "insulator" layer and sufficient suppression of upward band-bending is achieved for recipe #3. The GaN polarization-induced upward band-bending substantially reduces the HBT current gain from 30 (recipe #3) to 0.8 for the bare GaN (recipe #1). Due to the strong insulating effects of the $Al_2O_3$ layer, the current gain was less than 1 for recipes #5 and #7. Since the $Al_2O_3$ layer is not an insulator in our material system, it is highly likely that the simulation in this case has underestimated the performance of the HBT.

Considering that neither of the two simulations can provide accurate predictions of the HBT performance, the simulation results can be regarded as a rough performance range estimation (from the best scenario, Case I, to the worst, Case II) of the HBT. While an accurate modeling approach is needed to simulate the grafted material systems, device fabrication work using the grafting approach will be needed to experimentally verify the GaN-collector HBT prospective, along with characterization of the interface such as interface trap distribution in the bandgap [5] to help us better understand the grafted interface with UO.